\documentclass[12pt]{article}        
\usepackage{amstex}
\usepackage{amssymb}
\usepackage{epsfig}
\usepackage{cite}
\usepackage{epic}
\usepackage{fancybox}
\begin{document}
\begin{titlepage}

\normalsize
\begin{flushleft}
{\tt hep-ph/9610427}\\
October 1996
\end{flushleft}
\vspace{0.3cm}
\begin{center}
\Large
{\bf The Small-${\boldmath x}$ Evolution of Unpolarized} \\
\vspace{0.1cm}
{\bf and Polarized Structure Functions$^\ast$} \\
\vspace{0.4cm}
\large
J. Bl\"umlein, S. Riemersma \\
\vspace{0.3cm}
\large 
{\it DESY--Zeuthen \\
\vspace{0.1cm}
Platanenallee 6, D--15735 Zeuthen, Germany }\\
\vspace{0.4cm}
\large
A. Vogt\\
\vspace{0.4cm}
\large 
{\it Institut f\"ur Theoretische Physik, Universit\"at W\"urzburg \\
\vspace{0.1cm}
Am Hubland, D--97074 W\"urzburg, Germany} \\
\vspace{1cm}
{\bf Abstract}
\end{center}
\vspace{-0.1cm}
\normalsize
A brief overview is presented of recent developments concerning resummed
small-$x$ evolution, based upon the renormalization group equation.
The non-singlet and singlet structure functions are discussed for both
polarized and unpolarized deep-inelastic scattering.  Quantitative
results are displayed and uncertainties from uncalculated subleading
terms are discussed.
\vfill 
\noindent
\normalsize
$^{\ast} $ Based on an invited talk presented by 
S. Riemersma at the {\sf Third International Symposium on Radiative
  Corrections (CRAD'96)\/}, Cracow, August~1996.  To appear in {\sf Acta
Physica Polonica}. 
 
\end{titlepage}
\begin{center}
{\large \bf The Small-{\boldmath $x$} Evolution of Unpolarized and 
\\
\vspace{0.2cm}
Polarized Structure Functions\footnote{Presented by S. Riemersma}}
\vskip0.5cm
J. Bl\"umlein, S. Riemersma,\\
\vspace{0.2cm}
DESY-Zeuthen\\
Platanenallee 6, D-15735 Zeuthen, Germany\\
\vspace{0.15cm}
and\\
\vspace{0.15cm}
A. Vogt\\
\vspace{0.2cm}
Institut f\"ur Theoretische Physik, Universit\"at W\"urzburg\\
Am Hubland, D-97074 W\"urzburg, Germany\\
\end{center}
\vspace{0.4cm}
{\small
A brief overview is presented of recent developments concerning resummed
small-$x$ evolution, based upon the renormalization group equation.
The non-singlet and singlet structure functions are discussed for both
polarized and unpolarized deep-inelastic scattering.  Quantitative
results are displayed and uncertainties from uncalculated subleading
terms are discussed.}

\vspace{0.5cm}

\centerline{\bf 1. Introduction}

\vspace{0.5cm}

\noindent
Many investigations have been performed analyzing the small-$x$
behaviour of deep-inelastic scattering (DIS) structure functions.  The
reasoning is the evolution kernels of the non-singlet and singlet,
polarized and 
unpolarized parton densities contain large logarithmic contributions
in the small-$x$ region and large effects are in principle observable
in colliders such as HERA.

The resummation of these small-$x$ terms to all orders in the strong
coupling $\alpha_s$ can be completely handled within the framework of
perturbative QCD.  As collinear and ultraviolet divergences
appear in the calculations, the only appropriate method of
incorporating the effects of the small-$x$ resummations is to use the
renormalization group equations.  Evolving in $Q^2$ is the only way
the effects of the resummation can be studied.

Additionally, the effects of the resummed anomalous dimensions upon
the DIS structure functions are strongly dependent upon the parton
densities at the initial $Q_0^2$, which are non-perturbative and must
be given as input.  As the evolution is based upon the Mellin
convolution of the parton 
densities with the anomalous dimensions, the large and medium-$x$
regions are taken into account as well.  The large logarithmic
contributions to the anomalous dimensions therefore do not
automatically imply a large effect upon the observable DIS structure
functions.

In section 2, the underlying principles of the all-order small-$x$
resummation based upon the renormalization group are recalled.
Section 3 presents numerical results\footnote{For a much more
complete and detailed discussion see [1].}.  The effect of
physically motivated subleading terms is also discussed, as well as additional
uncertainties in cases where the input parton densities are not well
constrained.  Section 4 contains the conclusions.

\vspace{0.5cm}

\centerline{\bf 2. Theoretical Background}

\vspace{0.5cm}
\noindent
The evolution of the parton densities is given, for the non-singlet case
(and generically for the singlet), by
\begin{equation}
 \frac{\partial\, q(x,Q^2)}{\partial \ln  Q^2} = P(x, \alpha_s)
 \otimes q(x,Q^2) ,
\end{equation}
where the $\otimes$ denotes the Mellin convolution.
This convolution is not simply limited to the low-$x$ region, rather
it is dependent upon the entire possible $x$-range.

The leading gluonic contributions to the unpolarized singlet anomalous
dimension behave according to [2] ($a_s \equiv \alpha_s(Q^2)/4 \pi$) as
\begin{eqnarray}
\bigg ( \frac{a_s}{N-1} \bigg )^k \leftrightarrow 
\frac{1}{x} a_s^k \ln^{k-1} x \: .
\end{eqnarray}
The corresponding quark anomalous dimensions, being one power down in
$\ln x$ have been calculated in [3].  The leading terms of all
anomalous dimensions for the non-singlet [4] and polarized singlet [5]
evolutions are given by
\begin{equation}
N \bigg ( \frac{a_s}{N^2} \bigg )^k \leftrightarrow a_s^k
\ln^{2k - 2} x \: .
\end{equation}
The splitting functions $P(x,a_s)$ can be represented by
\begin{equation}
P(x,a_s) = \sum^{\infty}_{l = 0} a_s^{l+1} P_l(x) \: .
\end{equation}
Order by order in $a_s$, the expansion coefficients $P_l(x)$ are
subject to the constraints
\begin{equation}
\int_0^1 \! dx \, P_{l}^-(x) = 0 \: , \: \int_0^1 \! dx \, x \sum_i
P_{ij,l}^{\rm unpol.}(x) = 0 \: , 
\end{equation}
where the first equation represents fermion number conservation
for the `--' non-singlet expansion
coefficients and the second equation four-momentum conservation in the
unpolarized singlet case.   The resummed anomalous dimensions are subject
to the existence of subleading terms and the
effects of such terms have been investigated in [1,6-10].  Physically
motivated examples
used here will be
\begin{eqnarray}
\label{eq:constraints}
A: \Gamma(N,a_s) &\rightarrow& \Gamma(N,a_s) - \Gamma(1,a_s) \nonumber \\
B: \Gamma(N,a_s) &\rightarrow& \Gamma(N,a_s)(1 - N) \nonumber \\
C: \Gamma(N,a_s) &\rightarrow& \Gamma(N,a_s)(1 - 2N + N^2) \nonumber \\
D: \Gamma(N,a_s) &\rightarrow& \Gamma(N,a_s)(1 - 2N + N^3) \: ,
\end{eqnarray}
where $N \rightarrow N-1$ for the unpolarized singlet case.  

Beyond leading-order, the parton densities themselves are not 
observables.  The parton densities must be Mellin convoluted with the
appropriate coefficient functions to engender the observable structure
functions.  The accumulated  effect upon the structure function
determines the real impact of the small-$x$ resummation.  The
effect of the yet uncalculated subleading terms may be illustrated by
the variance 
of the prescriptions $A-D$.  Only when the variance is small and the
results are similar to the original resummation-enhanced structure
function can the resummation be considered reliable.  

\vspace{0.5cm}

\centerline{\bf 3. Numerical Results}

\vspace{0.3cm}
\noindent
{\it 3.1 Non-singlet structure functions}

\vspace{0.3cm}
\noindent
The evolution of the `--'-combination
\begin{eqnarray}
\label{F3N}
 \lefteqn{ xF_{3}^{\, N}(x,Q_0^2) \equiv \frac{1}{2} \!\left[ xF_{3}^{\,
    \nu N} (x,Q_0^2) + xF_{3}^{\, \bar{\nu} N}(x,Q_0^2) \right ] }
    \nonumber \\
 & & = c_{F_3}^-(x,Q^2_0) \otimes [xu_v + xd_v](x,Q_0^2)
\end{eqnarray}
for an isoscalar target $N$ and the  `+'-combination
\begin{eqnarray}
\label{F2pn}
  \lefteqn{ F_{2}^{\, ep}(x,Q_0^2) - F_{2}^{\, en}(x,Q_0^2)
  = } \\
 & & \hspace{-7mm} c_{F_2}^+(x,Q^2_0) \otimes \frac{1}{3}
  \!\left[ xu_v - xd_v - 2(x\bar{d}-x\bar{u})\right]\! (x,Q^2_0)
  \nonumber
\end{eqnarray}
have been investigated in [7,8]. As in all other
numerical examples displayed below, the reference scale for the
evolution is chosen as $Q_{0}^{2} = 4 \mbox{ GeV}^2$, and
the same input parameters are employed for the next-to-leading-order
(NLO) and the resummed 
calculations. In the present case, the initial parton distributions
have been  adopted from the MRS(A) global fit [11] together with 
the value of the QCD scale parameter, $\Lambda_{\overline{\rm MS}}
(N_f =4) = 230$ MeV.  From figure 1, the resummation effect can be
seen to be on the order of one percent and not giving a K-factor on
the order of ten as suggested in [12].

The effects of the small-$x$ resummation can also be considered in the
context of QED.  The effects have been investigated for initial-state
QED radiative corrections to deep-inelastic $eN$ scattering.  For
large $y$ and small $x$, the effect can reach up to ten percent [13].
The resummation was also investigated for $e^+ e^-
\rightarrow \mu^+ \mu^-$.  The results can also be found in [13].

The polarized non-singlet case presents interesting features in addition to
those observed in the unpolarized non-singlet.  In contrast to the
unpolarized situation, the shapes of the input parton densities have
not been well established yet.  An additional freedom is available to
adjust the input densities and gauge the impact.

\begin{center}
\vspace*{-20mm}
\mbox{\hspace*{-2mm}\epsfig{file=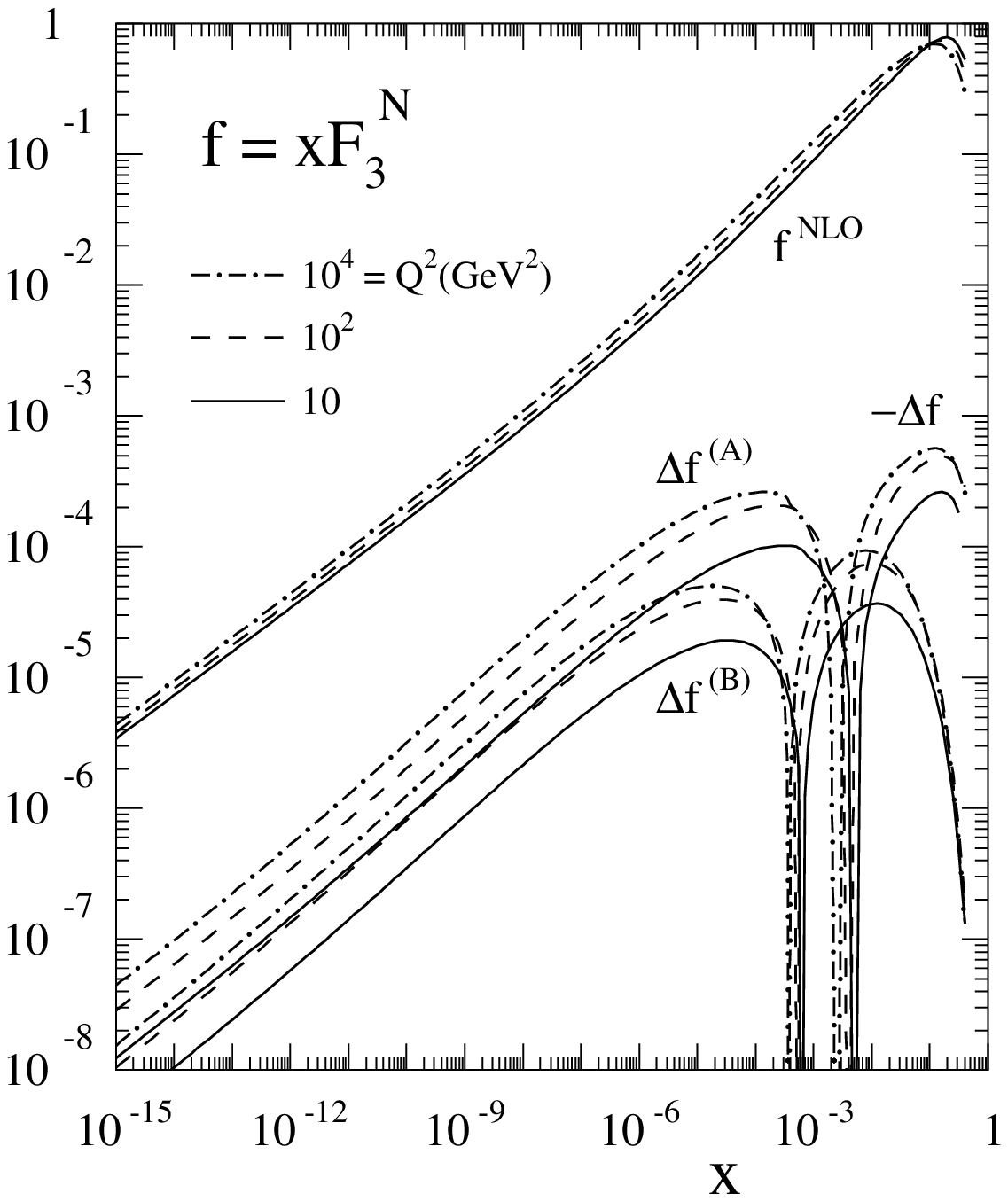,height=8cm,width=6.25cm}
      \hspace{-2mm}\epsfig{file=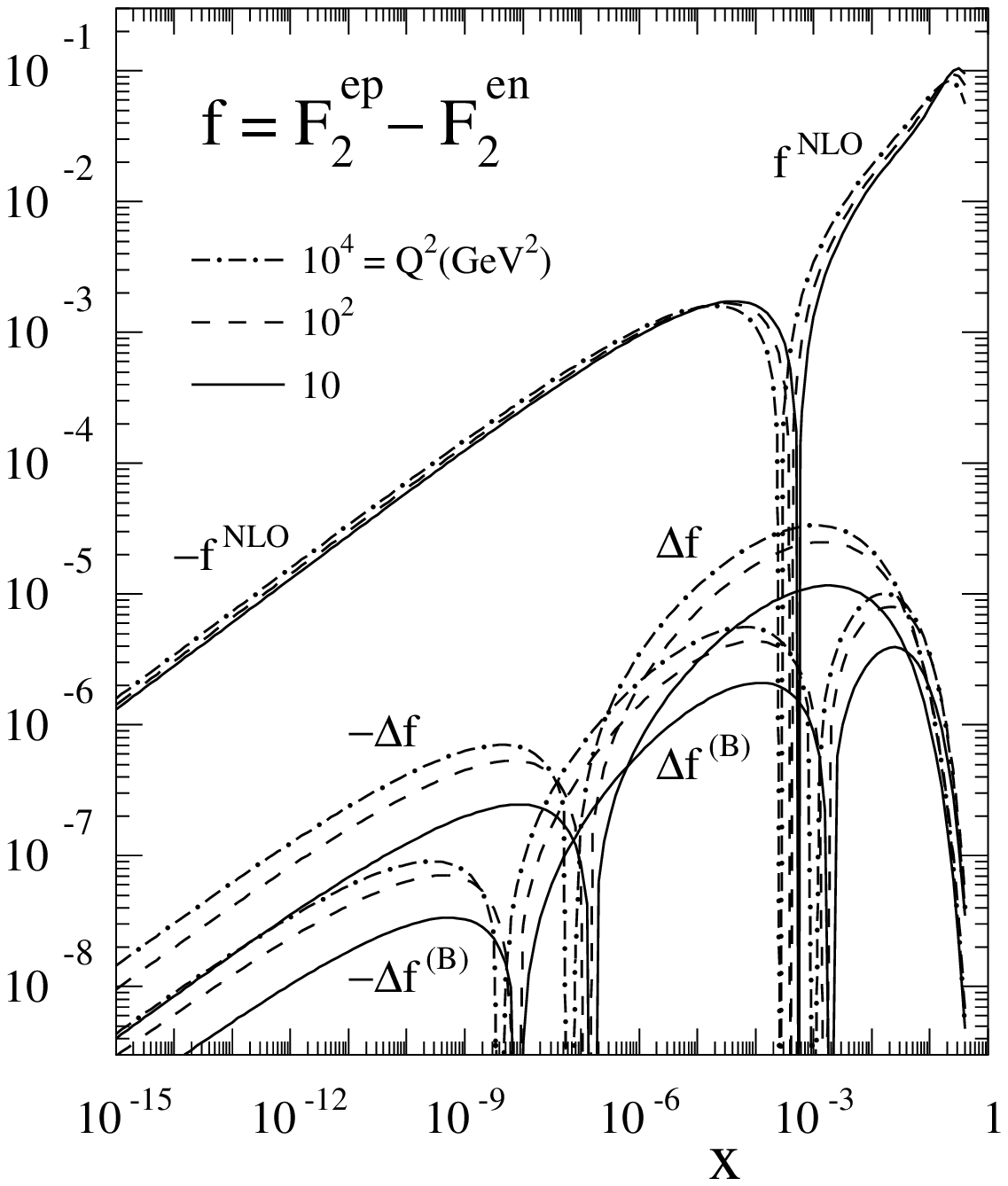,height=8cm,width=6.25cm}}
\vspace*{-6mm}
\end{center}
\vspace*{-3mm}
{\sf {\bf Figure 1a. (left):}~~The small-$x$ $Q^2$-evolution of
  $xF_3^{\, N}$ in NLO and the corrections to these results due to the
  resummed kernels. `(A)' and 
 `(B)' denote two prescriptions for implementing fermion 
 number conservation, see eq.~(\protect\ref{eq:constraints}).}
{\sf {\bf Figure 1b. (right)}~~The same as in Figure 1a., but for the structure
 function combination $F_2^{\, ep} - F_2^{\, en}$. Instead of the
 prescription `(A)', the result without any subleading terms is shown
 for this `+'-case.}

\vspace{0.3cm}

The effect on similar polarized non-singlet combinations
is an enhancement on the order of 15 \% at $x \sim 10^{-5}$ using the
fermion-number 
conserving prescription (A) but disappears completely when using (D)
for flat input parton densities.
For the steep densities, the effect is maximally 1.5 \%, and
can also be eliminated depending upon the choice of fermion-number
conservation prescription.

\vspace{0.3cm}
\noindent
{\it 3.2 Polarized singlet structure functions}

\vspace{0.3cm}
\noindent
Resummation relations for amplitudes [4] related to the singlet anomalous
dimensions for polarized DIS have been derived in [5].    Explicit
analytic and numerical results for the evolution kernels beyond NLO
have been derived using these relations in [6], including an all-order
symmetry relation among the elements of the anomalous dimension matrix
and a discussion of the supersymmetric case.

Here also we suffer from poorly constrained input parton densities,
adding to the uncertainty surrounding the effects of the resummation.

\begin{center}
\vspace*{-5mm}
\mbox{\hspace*{-2mm}\epsfig{file=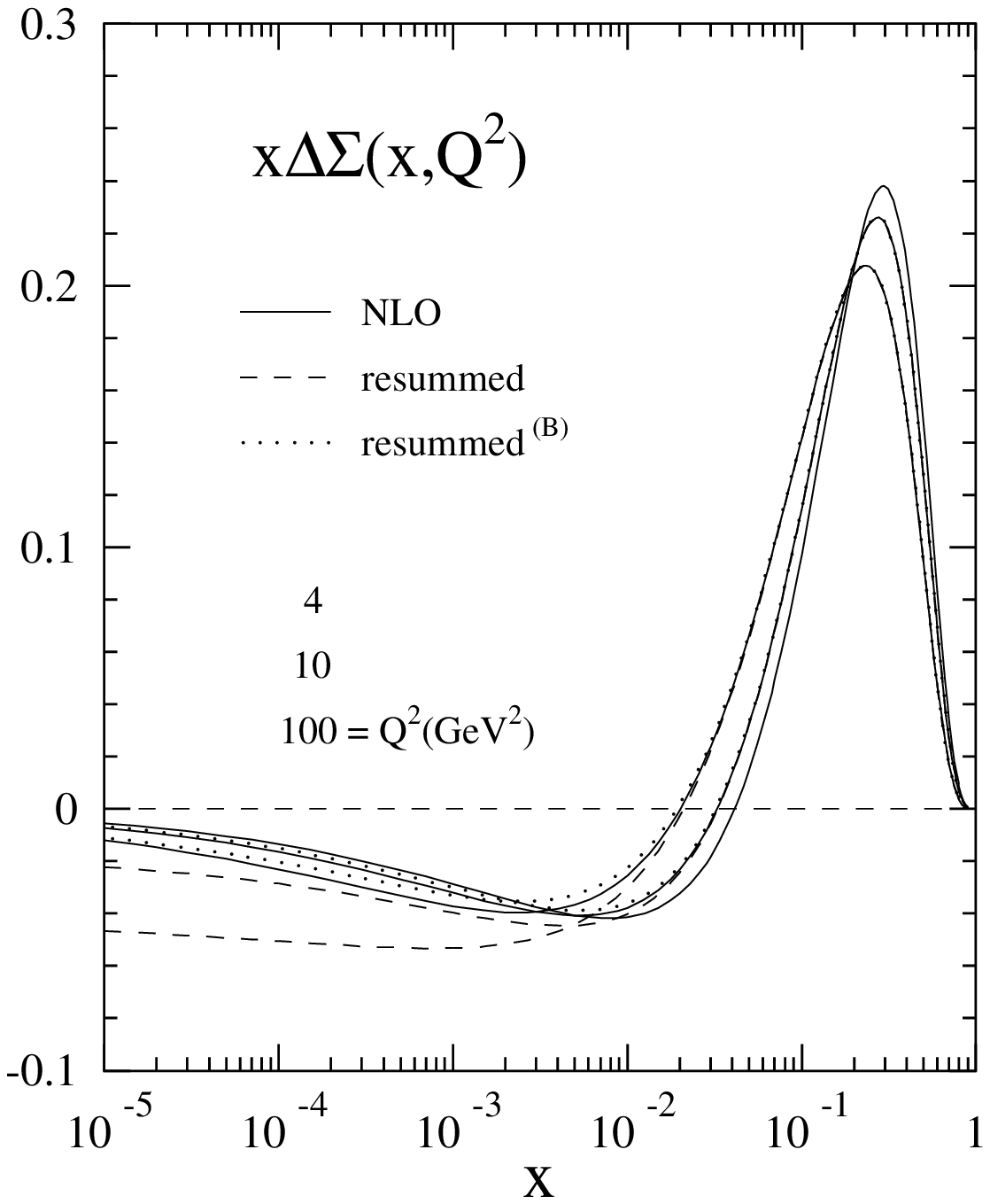,height=6.25cm,width=6.25cm}
      \hspace{-2mm}\epsfig{file=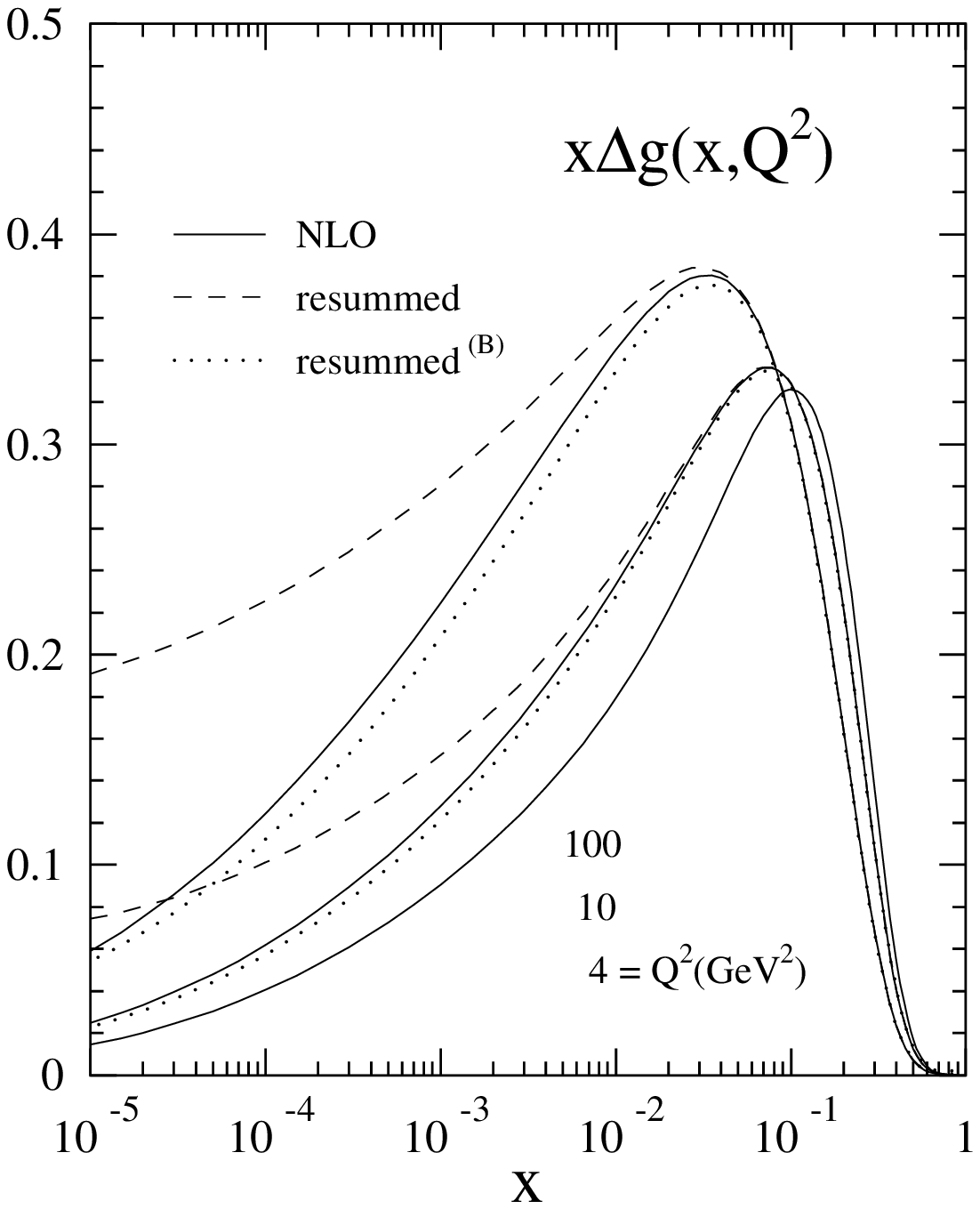,height=6.25cm,width=6.25cm}}
\vspace*{-6mm}
\end{center}
{\sf {\bf Figure 2a (left):}~~The evolution of the polarized singlet
  combination 
$x\Delta \Sigma$ in NLO and including the resummed kernels.  The impact 
of possible subleading terms is illustrated by the prescription `(B)' 
in eq.~(\ref{eq:constraints}). The input densities are from ref.~[14].}
{\sf {\bf Figure 2b (right):}~~As in Figure 2a, but for the polarized gluon
momentum distribution $x \Delta g $. As in the previous figure, the
$Q^2$-values in the legend are ordered according to the sequence of the
curves at small $x$.}

\vspace{0.3cm}

From figures 2a and 2b, the effects of the resummation are quite
evident.  The resummation has a considerable impact, in particular
upon $x \Delta g(x,Q^2)$.  While no sum rule is applicable to the
polarized singlet 
case, we draw upon the knowledge that the coefficients of the terms of
the anomalous dimensions subleading by one power in $N$ at LO and NLO
are generally of the same magnitude and of the opposite sign [6].  We
can therefore take this as an example of what subleading terms as $x
\rightarrow 0$ might be.  

\vspace{0.3cm}
\noindent
{\it 3.3 Unpolarized singlet structure functions}

\vspace{0.3cm}
\noindent
We now turn to the unpolarized singlet case, where the LO small-$x$
resummations  have been performed in [2] and the NLO quark sector
resummations by [3].  This case is of particular importance as
the high-statistics data from HERA are beginning to arrive, testing
the physical viability of the resummation as well as the NLO
calculations of DIS structure functions.  Investigations have been
carried out in [1,9,10] and an example is displayed in figure 3.  

\vspace{0.3cm}
\begin{center}
\vspace*{-5mm}
\mbox{\hspace*{-5mm}\epsfig{file=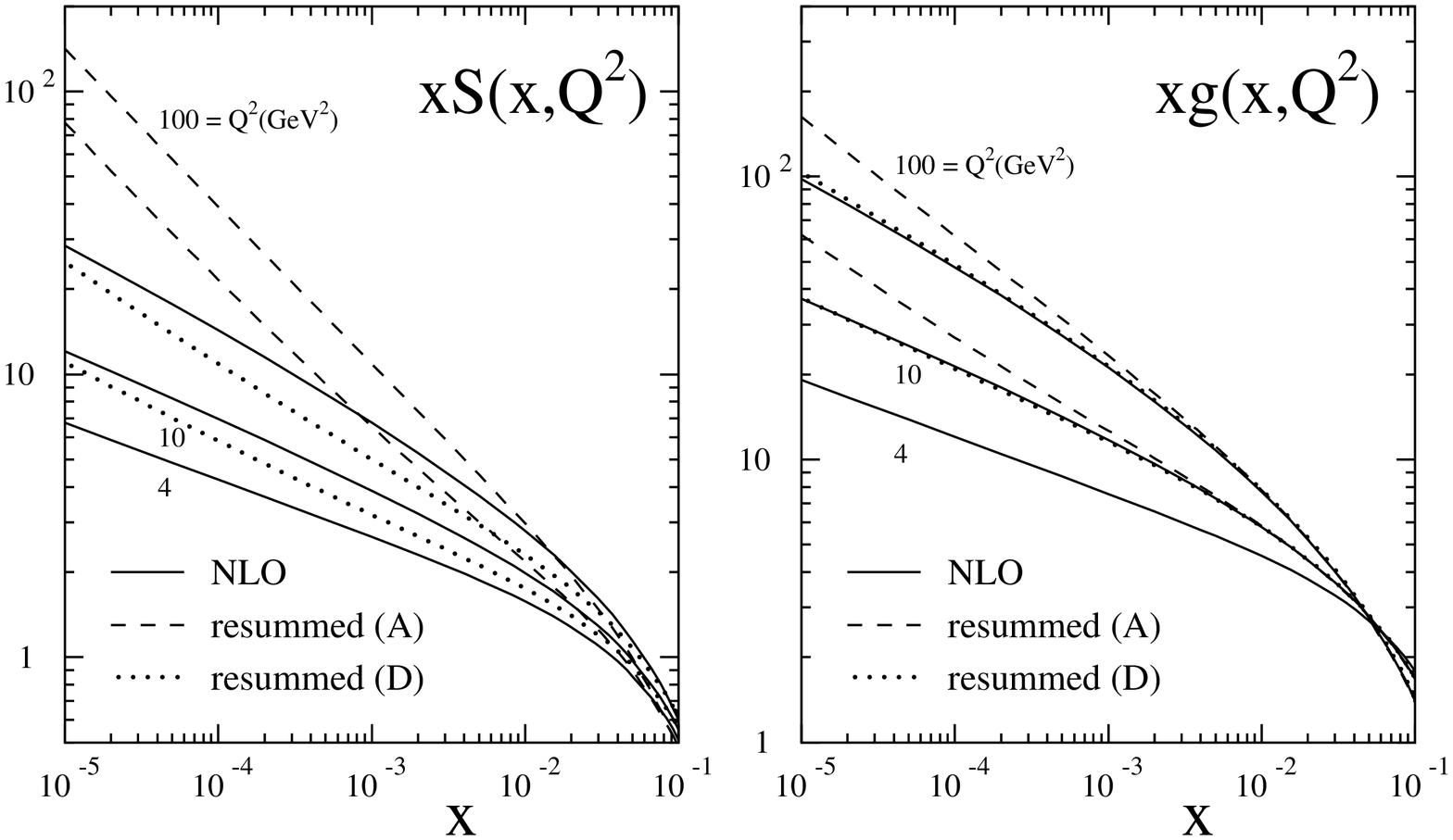,height=6.5cm,width=12cm}}
\vspace*{-6mm}
\end{center}
{\sf {\bf Figure 3:}~~The evolution of the gluon and total sea
  densities with the resummed kernels of [2,3] compared to the NLO
  results.  Prescriptions `(A)' and `(D)' have been implemented (see
  eqn. (\ref{eq:constraints})).}
\vspace{0.3cm}

As in the polarized singlet resummation, the effects of the pure
resummations are quite large.  The inclusion of the four-momentum
conserving subleading terms, however, substantially reduces the
effects and the resummed results can even fall below the NLO curve
using (D).

\vspace{0.5cm}

\centerline{\bf 4. Conclusions}

\vspace{0.5cm}
\noindent
The current status of small-$x$ resummations of polarized and
unpolarized, non-singlet and singlet structure functions has been
discussed.  

The unpolarized non-singlet structure functions are
enhanced by one percent or less.  Similar observations have been made
for the polarized non-singlet case.  The non-singlet QED corrections are
found to have an effect on the order of ten percent for $x \sim
10^{-4}$ and $y > 0.9$.

The singlet cases for both the polarized and unpolarized structure
functions show large effects stemming from the resummation.  Taking
into account less singular terms, arising from energy-momentum
conservation in the unpolarized situation or observing the behaviour
of the LO and NLO polarized anomalous dimensions, reduces the effect
considerably and can even completely eliminate it.

To assess the real effect of the small-$x$ resummations, the
subleading terms need to be calculated.  As demonstrated here, the
subleading terms may be quite important in the evolution of the
structure functions.  To have an adequate foundation for comparison,
the next-to-next-to-leading order results need to be calculated.

\vspace{0.5cm}
\noindent
{\bf Acknowledgments:}  This work was supported in part by the EC
Network `Human Capital and Mobility' under contract
No. CHRX-CT923-0004, and by the German Federal Ministry for Research
and Technology (BMBF) under contract No. 05 7WZ91P (0).

\vspace{0.5cm}

\centerline{\bf References}

\vspace{0.5cm}

\noindent
1. J. Bl\"umlein, S. Riemersma, and A. Vogt, {\sf hep-ph/9608470},
Nucl. Phys. {\bf B} (Proc. Suppl.) {\bf C51} (1996) 30.\\
\noindent
2. Y. Balitsky and L. Lipatov, Sov. J. Nucl. Phys. {\bf 28} (1978)
822.\\
\noindent
3. S. Catani and F. Hautmann, Nucl. Phys. {\bf B427} (1994) 475.\\
\noindent
4. R. Kirschner and L. Lipatov, Nucl. Phys. {\bf B213} (1983) 122.\\
\noindent
5. J. Bartels, B. Ermolaev, and M. Ryskin, DESY 96-025.\\
\noindent
6. J. Bl\"umlein and A. Vogt, Phys. Lett. {\bf B386} (1996) 350.\\
\noindent
7. J. Bl\"umlein and A. Vogt, Phys. Lett. {\bf B370} (1996) 149.\\
\noindent
8. J. Bl\"umlein and A. Vogt, Acta Physica Polonica {\bf B27} (1996) 1309.\\
\noindent
9. J. Bl\"umlein, S. Riemersma, and A. Vogt, {\sf hep-ph/9607329},{\it
  Proceedings of the International Workshop on 
  Deep Inelastic Scattering and Related Phenomena, Rome, April 1996},
to appear.\\
10. R.K. Ellis, F. Hautmann, and B. Webber, Phys. Lett. {\bf B348}
(1995) 582.\\
\noindent
11. A.D. Martin, R. Roberts, and J. Stirling, Phys. Rev. {\bf D50}
(1994) 6734.\\
\noindent
12. B. Ermolaev, S. Manayenkov, and M. Ryskin, Z. Physik {\bf C69}
(1996) 259.\\
\noindent
13. J. Bl\"umlein, S. Riemersma, and A. Vogt, DESY 96-120.\\
\noindent
14. M. Gl\"uck, E. Reya, M. Stratmann, and W. Vogelsang,
Phys. Rev. {\bf D53} (1996) 4775.
\end{document}